\documentclass[aps,prx,twocolumn,superscriptaddress,showpacs]{revtex4-1}
\usepackage{amsmath,amssymb,graphics,epsfig,epstopdf,color,verbatim,ulem,braket,tabularx}
\usepackage{multirow}
\usepackage[colorlinks,linkcolor=blue,citecolor=blue,urlcolor=blue,bookmarks=false]{hyperref}

\usepackage{listings}
\usepackage{cancel}
\usepackage{mathrsfs}
\usepackage{soul}
\usepackage{color}

\begin{document}

\title{Unconventional Pairing from Local Orbital Fluctuations in Strongly Correlated A$_3$C$_{60}$}

\author{Changming Yue}
\email{changming.yue@unifr.ch}
\affiliation{Department of Physics, University of Fribourg, 1700 Fribourg, Switzerland}

\author{Shintaro Hoshino}
\affiliation{Department of Physics, Saitama University, Saitama 338-8570, Japan}

\author{Akihisa Koga}
\affiliation{Department of Physics, Tokyo Institute of Technology, Meguro, Tokyo 152-8551, Japan}

\author{Philipp Werner}
\email{philipp.werner@unifr.ch}
\affiliation{Department of Physics, University of Fribourg, 1700 Fribourg, Switzerland}

\begin{abstract}
The pairing mechanism in A$_3$C$_{60}$ is investigated by studying the properties of 
a three-orbital Hubbard model with antiferromagnetic Hund coupling in the normal and superconducting phase.
Local orbital fluctuations are shown to be substantially enhanced in the superconducting state, with a fluctuation energy scale that matches the low-energy peak in the spectral weight of the order parameter.  
Our results demonstrate that local orbital fluctuations provide the pairing glue in strongly correlated fulleride superconductors and support the spin/orbital freezing theory of unconventional superconductivity. 
They are also consistent with the experimentally observed universal relation between the gap energy and local susceptibility in a broad range of unconventional superconductors. 
\end{abstract}

\maketitle

The phenomenon of unconventional superconductivity in strongly correlated electron systems (SCES) remains mysterious and debated even decades after the discovery in different classes of materials, including heavy Fermion systems \cite{Aoki2019,Hoshino2015}, cuprates \cite{Anderson2007,Maier2008,Werner2016,Sakai2016}, strontium ruthenate \cite{Rice1995,Hoshino2015,Pustogov2019} and fulleride compounds \cite{Capone2009,Hoshino2017}. While most researchers agree that the mechanisms in these materials are different from conventional phonon-mediated pairing, and more likely related for example to spin fluctuations \cite{Scalapino2012}, it is difficult to provide convincing evidence for a given scenario because of the challenges of analytical and numerical treatments of SCES. Nevertheless there has been important progress on the theoretical side in recent years. Model calculations on the strongly interacting repulsive and attractive Hubbard model have revealed peculiar cancellations between spectral signatures which have been interpreted as evidence for a coupling to a ``hidden Fermion" \cite{Sakai2016}. The clarification of the origin and nature of this fermionic excitation may result in a deeper understanding of unconventional superconductivity \cite{Imada2019}. In a separate effort, it was shown that superconductivity in strongly correlated multi-orbital Hubbard systems is closely linked to enhanced spin \cite{Hoshino2015} or orbital \cite{Steiner2016,Hoshino2017} fluctuations and the phenomenon of spin/orbital freezing \cite{Werner2008}. This observation also applies to the two-dimensional Hubbard model, where an auxiliary multi-orbital problem can be constructed from cluster orbitals \cite{Werner2016}. While the spin/orbital freezing theory of unconventional superconductivity is applicable to different classes of materials, from uranium based superconductors to cuprates, to fulleride compounds, and in this sense is universal \cite{Werner2016}, it has so far never been tested inside the superconducting phase. In this study, we provide the direct link between local moment fluctuations and unconventional superconductivity in strongly interacting electron systems. 

As target system, we consider the three-orbital Hubbard model with antiferromagnetic Hund coupling, which captures the physics of the fulleride superconductors A$_3$C$_{60}$ \cite{Capone2009}. These are three-dimensional, strongly correlated materials with three half-filled bands of $t_{1u}$ symmetry, which are clearly separated from the other bands \cite{Nomura2012}. Due to the extended molecular orbitals, the bare Hund coupling is small, so that the coupling to Jahn-Teller phonons can invert the sign of the effective static Hund coupling \cite{Fabrizio1997,Capone2002,Nomura2015}. This is the origin of the unconventional properties, which have been studied extensively \cite{Capone2009,Nomura2015,Hoshino2017} using dynamical mean field theory (DMFT) \cite{Georges1996}. DMFT is an appropriate method because of the large connectivity of the C$_{60}$ molecules in A$_3$C$_{60}$ and the importance of local (Hund) physics. For simplicity, we will consider a three-orbital Hubbard model with density-density interactions 
\begin{align}
&H_\text{loc}=\sum_{\alpha} U n_{\alpha\uparrow}n_{\alpha\downarrow} \nonumber\\
&\quad + \sum_{\sigma,\alpha>\gamma}[(U-2J) n_{\alpha\sigma}n_{\gamma\bar\sigma}+(U-3J)n_{\alpha\sigma}n_{\gamma\sigma}]
\end{align}  
and orbital-diagonal hopping on a Bethe lattice with bare bandwidth $2D$. Here $U$ is the intra-orbital repulsion, $J$ the effective Hund coupling, $\alpha,\gamma=1,2,3$ denotes the orbitals and $\sigma$ the spin. This set-up allows efficient DMFT simulations based on the segment representation \cite{Werner2006} of the hybridization expansion impurity solver \cite{Gull2011}. To treat the superconducting (SC) state, we use a Nambu implementation as described in Refs.~\cite{Georges1996,Koga2015}. The Green's functions $\hat G$ and self-energies $\hat \Sigma$ are orbital-diagonal, with a $2\times 2$ matrix structure:
\begin{equation}
\hat G_\alpha (\tau) = \left(
\begin{array}{cc}
G^\text{nor}_{\alpha\uparrow}(\tau) & F_\alpha(\tau)\\
F^*_\alpha(\tau) & -G^\text{nor}_{\alpha\downarrow}(-\tau)
\end{array}
\right)
\end{equation}
and similarly for $\hat \Sigma$. $G^\text{nor}_{\alpha\sigma}(\tau)=-\langle \mathcal{T} c_{\alpha\sigma}(\tau)c^{\dagger}_{\alpha\sigma}\rangle$ and $F_{\alpha}(\tau)=-\langle \mathcal{T} c_{\alpha\uparrow}(\tau)c_{\alpha\downarrow}\rangle$ denote the normal and anomalous Green's functions for orbital $\alpha$. 
The Hund coupling will be set to $J=-U/4$, which is larger than in the realistic compounds \cite{Nomura2012,Nomura2015}, but does not change the physics at a qualitative level \cite{Hoshino2017}. From the numerical point of view the advantage is that superconductivity appears at higher temperatures ($T$).  Some results for smaller $J$ are provided in the Supplemental Material (SM). 

\begin{figure}[t]
\includegraphics[clip,width=3.in,angle=0]{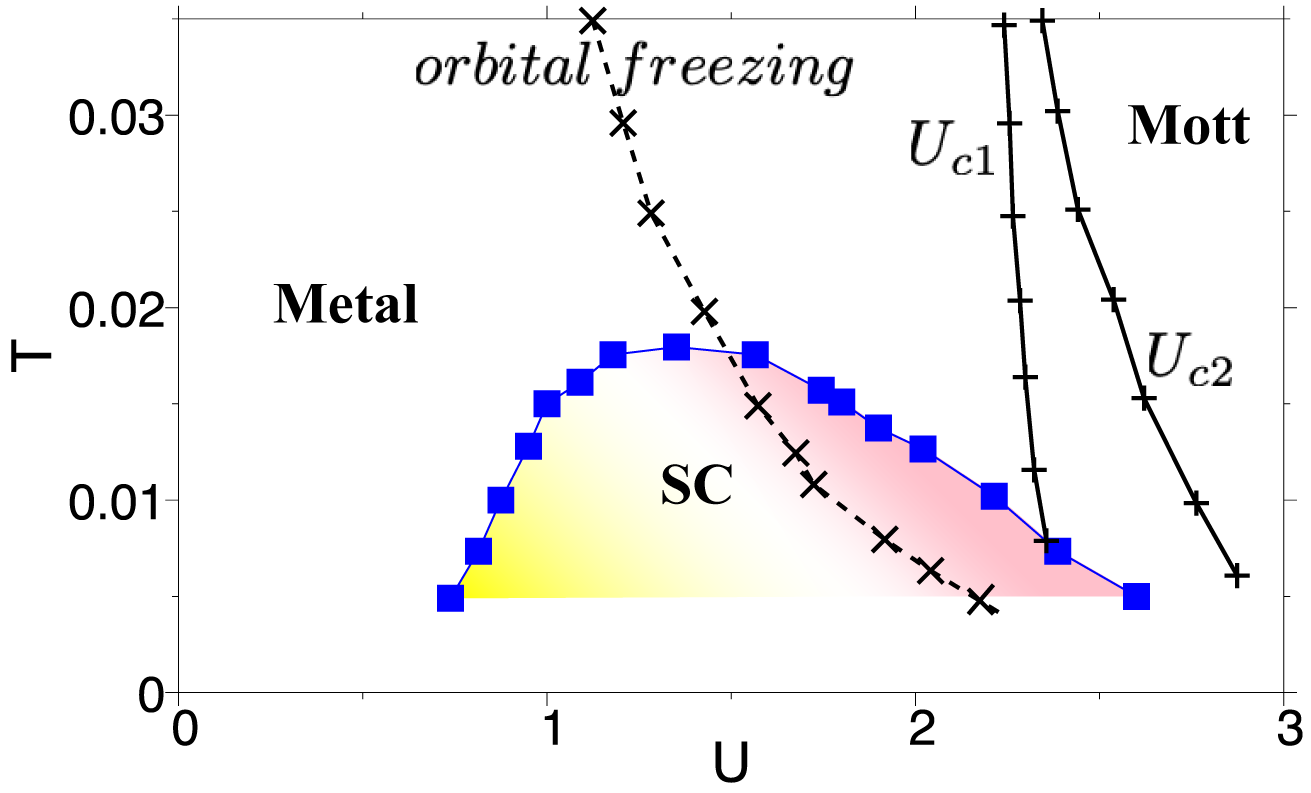}
\caption{(color online) Phase diagram of the three-orbital Hubbard model with $J=-U/4$ (reproduced from 
Refs.~\cite{Ishigaki2018,Hoshino2017}). The blue line shows the SC dome and we use a yellow (pink) shading to indicate the weak-coupling (strong-coupling) regime. 
The solid black lines indicate the $U_{c1}$ and $U_{c2}$ lines of the Mott transition and the dashed line locates the orbital freezing crossover in the normal phase.
}
\label{fig:phasediagram}
\end{figure}

Figure~\ref{fig:phasediagram} shows the DMFT phase diagram of the half-filled model as established in Refs.~\onlinecite{Hoshino2017,Ishigaki2018}. It features a SC dome as a function of $U$, which borders a Mott insulating phase. Not shown is the spontaneous orbital-selective Mott \cite{Hoshino2017,Ishigaki2018} (or Jahn-Teller metal \cite{Zadik2015}) phase between the SC dome and the Mott region, since this type of symmetry breaking, as well as orbital and magnetic orders \cite{Ishigaki2019}, will be suppressed in the following analysis, which focuses on the SC state. The phase diagram is qualitatively consistent with that of A$_3$C$_{60}$ \cite{Zadik2015}, where the SC dome and Mott transition line has been mapped out by the application of chemical pressure. 
As in the case of other unconventional SCs, the dome shape indicates a crossover from a weak-coupling to a strong-coupling regime. 
For example, K$_3$C$_{60}$ and Rb$_3$C$_{60}$ (located on the weak-$U$ side of the dome) are sometimes treated as ``conventional" SCs \cite{Zadik2015,Mitrano2016,Kasahara2017}, while overexpanded Rb$_x$Cs$_{3-x}$C$_{60}$ (on the strong-$U$ side of the dome) is an unconventional SC which clearly violates the BCS prediction for the ratio between the SC gap and $T_c$ \cite{Zadik2015}. In Fig.~\ref{fig:phasediagram} we indicate these two regimes with the yellow and pink shading, and we will use this color code also in the following figures.   

A relevant insight from recent DMFT studies \cite{Steiner2016,Hoshino2017} is that the superconductivity in multi-orbital Hubbard models with negative $J$ is intricately linked to an orbital freezing crossover. This is suggested by the fact that the SC dome peaks in the region of the phase diagram where the local orbital fluctuations, measured by the quantity $\Delta\chi_{\text{loc}}^\text{orb}=\int_0^\beta d\tau[\langle O_i(\tau)O_i(0)\rangle-\langle O_i(\beta/2)O_i(0)\rangle]$ (with $O_i$ some appropriately defined orbital moment), reach a maximum in the normal phase (see dashed line in Fig.~\ref{fig:phasediagram}). Using a weak-coupling picture, the enhancement in this quantity can be related to an effective attraction \cite{Hoshino2017} 
\begin{equation}
U_\text{eff} \sim U-4U'(U'+|J|)\Delta\chi_\text{loc} + O(U^3),
\label{eq:weak_coupling}
\end{equation}
and hence (indirectly) to SC. 
Here, we will provide direct evidence which links orbital fluctuations with $O_i=(n_1+n_2-2n_3)/\sqrt{3}$ in the present three-orbital model to SC. 

\begin{figure}[t]
\includegraphics[clip,width=3.4in,angle=0]{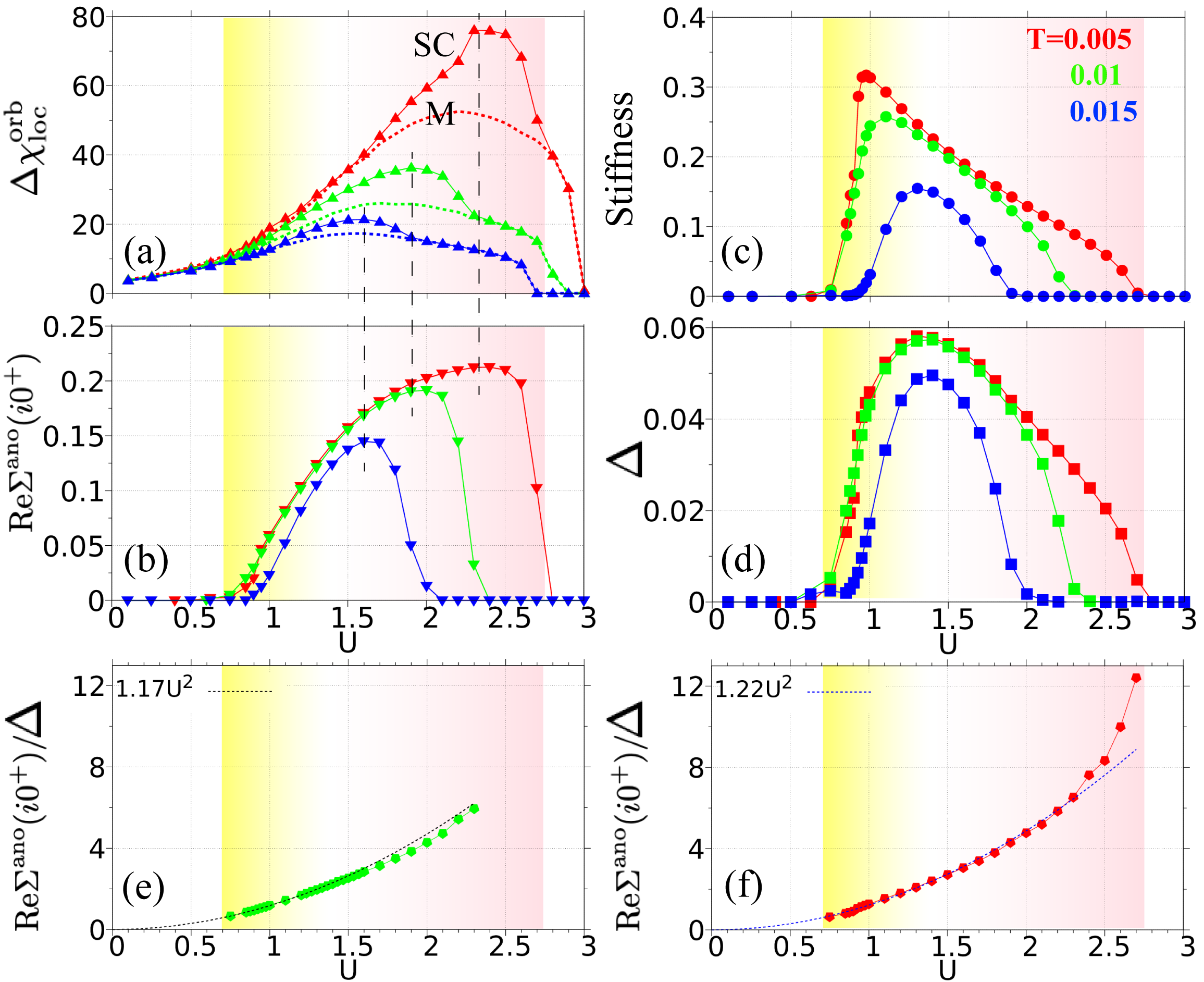}
\caption{ $U$-dependence of the local orbital fluctuation $\Delta\chi^{\text{orb}}_{\text{loc}}$, $\mathrm{Re}\Sigma^{\text{ano}}(i0^+)$,
          stiffness, order parameter $\Delta$ for $T=$0.005 (red), 0.01 (green), and 0.015 (blue), respectively. The dashed lines in panel (a) show the 
	 fluctuations in the normal metal phase. Panel (e) and (f) shows the ratio of $\mathrm{Re}\Sigma^{\text{ano}}(i0^+)/\Delta$
         at $T=0.01$ and $T=0.005$, respectively, and the black-dashed lines a quadratic fit to the data points.
         }
\label{fig:stiff_pairpot_chi_s12wn0}
\end{figure}

We start by comparing $\Delta\chi^\text{orb}_{\text{loc}}$ to other SC related quantities which can be evaluated directly on the Matsubara axis (see Fig.~\ref{fig:stiff_pairpot_chi_s12wn0}). The first important observation is that $\Delta\chi_{\text{loc}}^\text{orb}$ is {\it enhanced} in the SC phase compared to the normal phase, i.e., the orbital fluctuations which have previously been suggested to play a role in the pairing \cite{Hoshino2015} are present and even stronger in the SC state. This becomes clear from the comparison between the solid (dashed) lines in panel (a), which show $\Delta\chi_{\text{loc}}^\text{orb}$ from simulations with (without) symmetry breaking.  The enhancement is particularly pronounced on the orbital-frozen side of the dome, where the transition into the SC state ``unfreezes" the local moments (see also SM).  
Hence, the peak in $\Delta\chi^\text{orb}_{\text{loc}}$ appears on the strong-coupling side and coincides with the maximum in the static anomalous self-energy $\text{Re}\Sigma^\text{ano}(i\omega_n\rightarrow i0^+)$ (panel (b)). This quantity measures the strength of the pairing 
\cite{Maier2008,Sakai2016}, so that the data are consistent with the picture of an unconventional pairing induced by orbital fluctuations, which is strong on the large-$U$ side of the dome. Furthermore, 
from the self-energy and the SC order parameter $\Delta=\langle c_{\alpha\uparrow} c_{\alpha\downarrow}\rangle=-F_\alpha(\tau=0^+)$, we may (see SM)
extract the effective interaction $U_\text{eff}^\Sigma=\text{Re}\Sigma^\text{ano}(i\omega_n\rightarrow i0^+)/\Delta$. As shown in Fig.~\ref{fig:stiff_pairpot_chi_s12wn0}(e,f), a quadratic dependence of $U_\text{eff}^\Sigma$ on the bare $U$, qualitatively consistent with Eq.~(\ref{eq:weak_coupling}), is found on the weak-coupling side of the SC dome, while an even stronger increase with $U$ is found on the strong-coupling side, especially at low $T$ (Fig.~\ref{fig:stiff_pairpot_chi_s12wn0}(f)) \footnote{Quantitatively the prefactor of the quadratic term in Eq.~(\ref{eq:weak_coupling}) does not match, which indicates that renormalized couplings must be considered to connect the numerical data to this effective weak-coupling description.}.
We also show in Fig.~\ref{fig:stiff_pairpot_chi_s12wn0}(c) the superfluid stiffness computed by the procedure described in Ref.~\cite{Toschi2005}. The stiffness peaks on the weak-coupling side of the dome, as in the case of the attractive Hubbard model \cite{Toschi2005}. The opposite tendencies in the stiffness and pairing strength as a function of $U$ lead to a dome in 
$\Delta$, 
which peaks somewhere in between the other two quantities, around $U=1.4D$. 

\begin{figure}[t]
\includegraphics[clip,width=3.4in,angle=0]{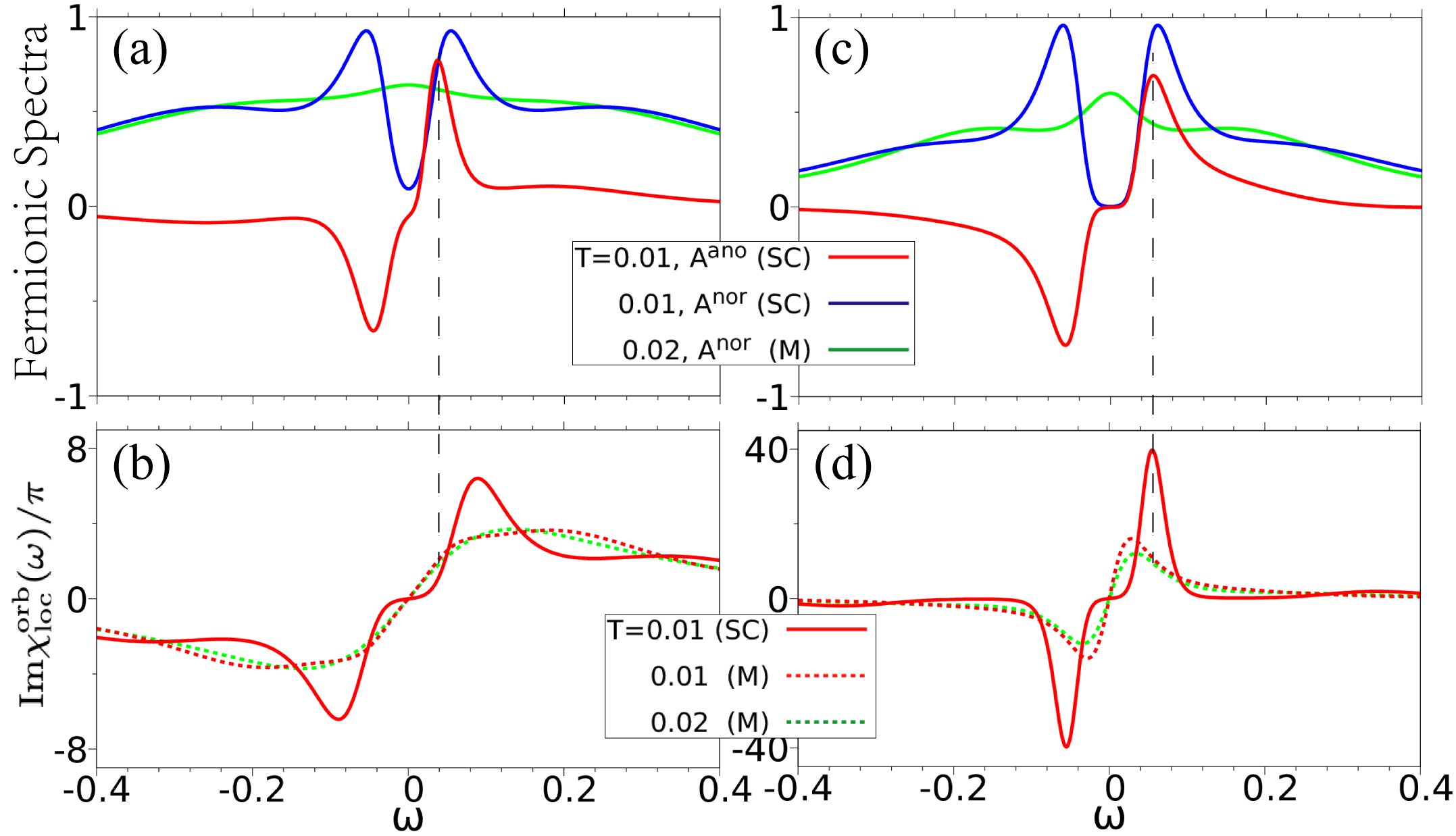}
\caption{(color online) Anomalous spectral function $A^{\mathrm{ano}}(\omega)$, 
normal spectral function $A^{\mathrm{nor}}(\omega)$ and bosonic spectrum $\tfrac1\pi\mathrm{Im}\chi^{\mathrm{orb}}_{\mathrm{loc}}(\omega)$.
Panels (a-b) are for weak coupling ($U=1.0$), and panels (c-d) for strong coupling ($U=1.9$).
The label M indicates the normal metal phase.
}\label{fig:Aw_Aanow_Chiw}
\end{figure}

\begin{figure}[t]
\includegraphics[clip,width=3.4in,angle=0]{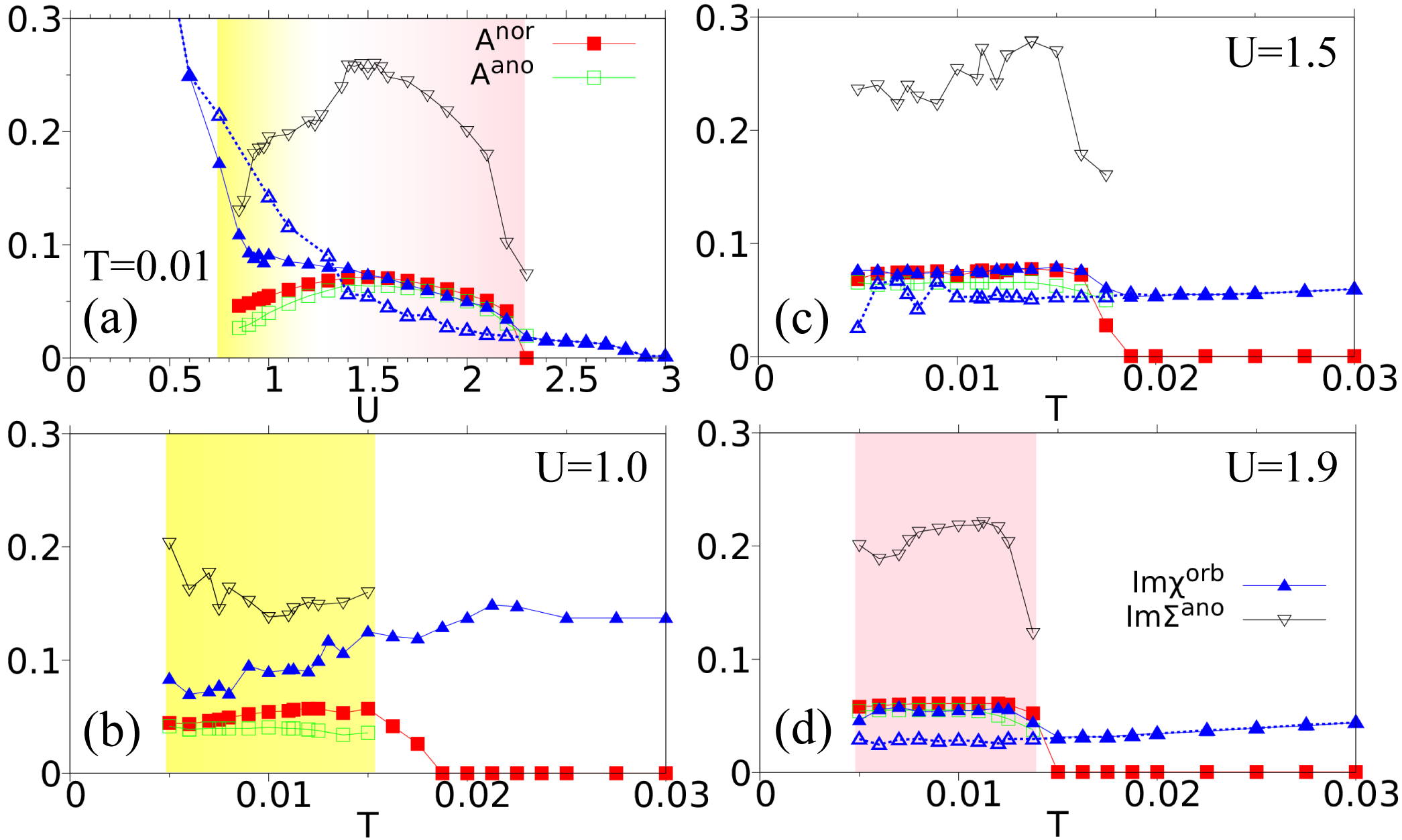}
\caption{(color online) Low-energy peak positions in $A^{\mathrm{nor}}(\omega)$,
$A^{\mathrm{ano}}(\omega)$,
$\mathrm{Im}\chi_{\text{loc}}^{\mathrm{orb}}(\omega)$,
and
$\mathrm{Im}\Sigma^{\mathrm{ano}}(\omega)$.
Panel (a) shows the $U$ dependence at $T=0.01D$, 
panels (b-d) show the $T$ dependence at (b) $U=1.0$, (c) $U=1.5$, and (d) $U=1.9$, respectively.
The blue dashed lines with empty triangles in panel (a) and (d) show the peak positions of $\mathrm{Im}\chi_{\text{loc}}^{\mathrm{orb}}(\omega)$
in the normal phase. The noise in the data is due to the limitations of MaxEnt analytical continuation.  
}
\label{fig:peaks_UxTx}
\end{figure}

Having revealed further evidence for the link between orbital fluctuations and SC in this fulleride-inspired model, we will now investigate real-frequency spectra. An important question concerns the characteristic energy scale of the local orbital fluctuations \cite{Maier2008} and their relation to the SC gap and peaks in the anomalous Green's function spectrum \cite{Kyung2009}. The spectra are computed with the Maximum Entropy method \cite{Bryan1990,Jarrell1996,LEVY2017149} with a bosonic or fermionic Kernel. 
In the case of the anomalous Green's function 
we employ the so-called MaxEnt-Aux method \cite{MaxEntAux}, where an auxiliary Green's function with positive spectral weight is introduced for the operator 
 $\hat{a}_{\alpha}=\frac{1}{\sqrt{2}}[c_{\alpha\uparrow}+c_{\alpha\downarrow}^{\dagger}]$ 
\cite{PhysRevB.91.085116}. In our particle-hole symmetric system, $G^{\mathrm{aux}}_{\alpha}(\tau)=G^{\mathrm{nor}}(\tau)+F(\tau)$ \cite{Gull_prl2013_SC_2D}. 
Representative spectra for the weak-$U$ and strong-$U$ side of the SC dome are shown in Fig.~\ref{fig:Aw_Aanow_Chiw}.
More results for different $U$ and $T$ can be found in the SM. In the normal phase, the bosonic spectrum $\tfrac1\pi\text{Im}\chi_{\text{loc}}^\text{orb}(\omega)$ associated with the local orbital fluctuations ($\chi_{\text{loc}}^\text{orb}(\tau)=\langle \mathcal{T}  O_i(\tau)O_i(0)\rangle$) exhibits a peak 
whose energy decreases with increasing $U$ (see blue dashed line with empty triangles in Fig.~\ref{fig:peaks_UxTx}(a))
\footnote{For simplicity of notation, we omit the retarded subscript in all spectral functions.}. It is however rather broad and in the large-$U$ regime there is spectral weight down to $\omega\rightarrow 0$, indicative of orbital freezing \cite{Steiner2016}. As temperature is lowered and the system enters into the SC phase, a gap in the bosonic spectrum opens and a sharp peak with an energy comparable to that in the normal phase appears (Fig.~\ref{fig:Aw_Aanow_Chiw}(b,d)). Looking at the fermionic spectra, we see the expected opening of a gap in the normal spectral function $A^\text{nor}(\omega)=-\tfrac1\pi\text{Im}G^\text{nor}(\omega)$ 
after the transition into the SC state, with sharp peaks near the gap edge. The anomalous spectral function $A^\text{ano}(\omega)=-\tfrac1\pi\text{Im}F(\omega)$ also exhibits a peak at a similar energy. Interestingly, on the strong-$U$ side of the SC dome, these peak positions are very close and  the energy of the bosonic peak matches that of the fermionic spectra 
almost exactly (Fig.~\ref{fig:Aw_Aanow_Chiw}(c,d)). Since $\int_0^\infty d\omega' A^\text{ano}(\omega')=\Delta$ at low $T$, the anomalous spectrum represents the spectral weight of the order parameter \cite{Kyung2009}. Hence, this match demonstrates a direct link between the orbital fluctuations and the pairing. 

Panel~\ref{fig:peaks_UxTx}(a) demonstrates that the SC gap exhibits a dome shape as a function of $U$, with a peak near the orbital-freezing crossover line, and that the almost perfect match between fermionic and bosonic spectral features starts around the maximum of that dome. 
In this crossover region, with increasing $U$, the orbital moments in the normal phase start to freeze and the orbital fluctuation energy scale in the normal phase (blue dashed line with empty triangles) drops below the energy scale of the order parameter (green line). Since the timescale of orbital fluctuations, which is given by the inverse of the peak energy in ${\rm Im}\chi^{\rm orb}_{\rm loc}$,  
cannot be longer than that of the relevant single particle fluctuations (roughly the inverse gap in the SC state), the peak in $\text{Im}\chi^{\text{orb}}_\text{loc}(\omega)$ cannot shift below the peak in $A^\text{nor/ano}(\omega)$ and the two fluctuation energy scales get locked in. This feedback of the SC state on the orbital fluctuations results in the melting of the orbital-frozen state, 
the release of the large entropy of the frozen metal state \cite{Yue2020}, an enhanced $\Delta\chi^{\text{orb}}_\text{loc}$, and (see Eq.~(\ref{eq:weak_coupling})) an enhanced pairing strength.

It is interesting to note in this context that an almost perfect match between the excitation energies in the single-particle and two-particle spectra is also found in Bose fluids \cite{Gavoret1965}, and it would be worthwhile to explore the possible connection to these systems in more detail.

The lock-in phenomenon is also clearly seen in panel (d), which plots the peak positions as a function of $T$ on the strong-coupling side of the SC dome. In the normal phase, the orbital fluctuation energy scale monotonously decreases with decreasing $T$ (blue dashed line with empty triangles), while the transition into the SC state leads to a synchronization with the (higher) single-particle excitation energy scale. 
In the orbital-freezing crossover regime (panel (c)), which features the largest gap, the results are qualitatively similar, but here the orbital fluctuation energy scale in the normal phase is already close to the synchronized orbital fluctuation and pairing energy scale in the symmetry-broken phase. On the weak-$U$ side of the dome, however, the lock-in phenomenon is absent (panel (b)), since the orbital moments in the normal phase are not frozen, and the orbital fluctuation energy scale is higher than the SC one, although it is reduced in the SC phase compared to the normal state. Here, we roughly find a factor of two between the peak position in $\tfrac1\pi\text{Im}\chi^\text{orb}_\text{loc}$ and $A^\text{nor/ano}$, as expected in the weak-correlation limit, where $\chi^\text{orb}_{\text{loc}}(\tau)=4[F(\tau)^2+G(\tau)^2]$.

Some previous studies of unconventional SCs have focused on the anomalous self-energy $\Sigma^\text{ano}$ instead of the anomalous Green's function $F$ \cite{Maier2008,Sakai2016}. Inspired by an analogy to phonon-mediated SCs, one may search for a peak in $\text{Im}\Sigma^\text{ano}(\omega)$, whose position corresponds to the sum of the gap edge energy ($\sim$ peak in $A^\text{nor}(\omega)$) and the bosonic excitation energy \cite{Maier2008}. 
For the calculation of $\text{Im}\Sigma^\text{ano}(\omega)$, we employ the procedure proposed in \cite{Gull_prl2013_SC_2D}, i.e., extract it from auxiliary self-energies $\Sigma_{\pm}(i\omega_n)=\Sigma^{\mathrm{nor}}(i\omega_n)\pm\Sigma^{\mathrm{ano}}(i\omega_n)$ with positive-definite spectral weight in the presence of particle-hole symmetry. 
The low-energy peak positions as a function of $T$ or $U$ are shown by the black lines in Figs.~\ref{fig:peaks_UxTx}(a) and \ref{fig:peaks_UxTx}(b-d). Clearly, this energy is higher than the above-mentioned sum on the large-$U$ side of the dome. Similarly, if we compute the ratio between the SC gap and $T_c$, the result is up to three times higher than the BCS prediction, indicating an unconventional pairing mechanism. 

\begin{figure}[t]
\includegraphics[clip,width=3.4in,angle=0]{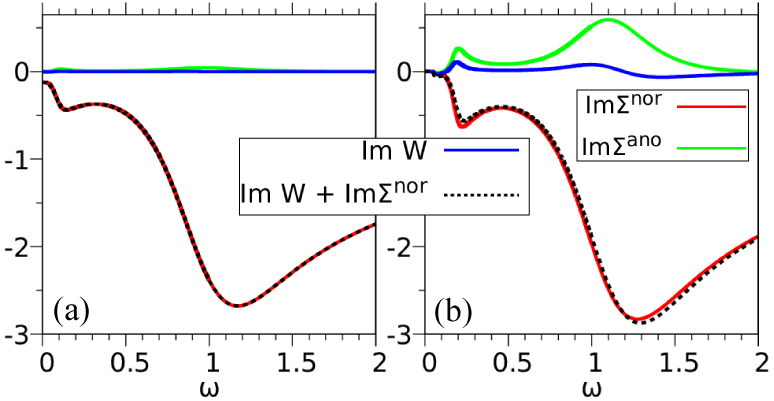}
\caption{(color online) 
Imaginary part of the real frequency normal and anomalous self-energy, and $\text{Im} W(\omega)$
for $T=0.015$ (panel (a)) and $T=0.005$ (panel (b)) at $U=1.9$. The dashed lines show 
$\text{Im} W(\omega)+\text{Im}\Sigma^{\text{nor}}(\omega)$.
}
\label{fig:Ww_Sw_Cancelation_U1p9}
\end{figure}

Finally, we would like to test the hidden Fermion scenario in our fulleride-SC-inspired model. According to Refs.~\cite{Sakai_prbR_2015,Sakai2016,Imada2019}, a remarkable property of the attractive Hubbard model and two-dimensional repulsive Hubbard model in the strongly correlated SC regime is an almost perfect cancellation between the peak in $\text{Im}\Sigma^\text{nor}(\omega)$ and a peak in $\text{Im}W(\epsilon=0,\omega)$, where $W(\epsilon=0,\omega)=\Sigma^\text{ano}(\omega)^2/[\omega-\mu+\Sigma^\text{nor}(-\omega)^*]$ \cite{Sakai_prbR_2015}. 
This implies that the sharp peak in $\text{Im}\Sigma^\text{nor}$ does not leave an obvious trace in $A^\text{nor}$. In Fig.~\ref{fig:Ww_Sw_Cancelation_U1p9} we plot $\text{Im}\Sigma^\text{nor}$, $\text{Im}W$ and $\text{Im}[\Sigma^\text{nor}+W]$ for $U=1.9$, at two temperatures close to and much below $T_c$. We find that the low-energy peaks have roughly the same positions and opposite signs, in agreement with the hidden Fermion prediction, but there is no cancellation. In fact, the peak in $\text{Im}\Sigma^\text{nor}$ is much more prominent than that in $\text{Im}W$. Also, a close inspection of the spectra reveals that $A^\text{nor}$ does in fact exhibit a shoulder structure associated with the peak in $\text{Im}\Sigma^\text{nor}$. Hence, the hidden Fermion mechanism seems not to be applicable to fulleride superconductors. 

Our findings are instead qualitatively similar to those for the square lattice repulsive Hubbard model reported in Ref.~\onlinecite{Kyung2009}. These authors demonstrated an analogous connection between the spectrum of local spin fluctuations and that of the order parameter, and a considerable mismatch between the peak in $\text{Im}\Sigma^\text{ano}$, shifted by the half-gap value, and the peak in the local spin fluctuation spectrum.  
This similarity suggests a universal pairing mechanism in unconventional SCs. According to the spin/orbital freezing theory of superconductivity \cite{Hoshino2015,Steiner2016,Werner2016,Hoshino2017}, the pairing mechanism in negative-$J$ and positive-$J$ multiorbital Hubbard systems is essentially the same, since the sign change of $J$ to a first approximation inverts the role of spins and orbitals \cite{Steiner2016}. Hence, orbital freezing induced SC is mapped to spin-freezing induced SC and vice versa, and we expect a completely analogous connection between local spin fluctuations and unconventional SC in  positive-$J$ systems. The mapping of the two-dimensional Hubbard model to an effective two-orbital model with $J>0$ and the spin-freezing related SC in this system has been discussed in Refs.~\cite{Werner2016,Werner2020}, and these results are fully consistent with the findings presented in our study and in Ref.~\onlinecite{Kyung2009}. 
With these facts in mind, it is very interesting to note that a universal linear relationship $E_r \approx 1.28\Delta$ between the magnetic resonance energy $E_r$ and the superconducting gap is experimentally found in a broad range of cuprates, iron pnictides and heavy electron materials \cite{NatPhys2009_Ereson_2SCgap}, which (see SM) appears to be the spin-freezing analog of the lock-in phenomenon revealed in this work.

{\it Acknowledgements} The calculations have been performed on the Beo05 cluster at the University of Fribourg, using a code based on ALPS \cite{alpa135} and iQist \cite{HUANG2015140,iqist}. We acknowledge support by SNSF Grant No. 200021-165539 and SNSF Grant No. 200021-196966.

\bibliography{Ref_singletSC.bib}

\clearpage
\newpage
\appendix
\numberwithin{equation}{section}
\numberwithin{figure}{section}

\section{Supplemental Material}

In this Supplemental Material we present additional data and analyses for the three-orbital model with $J<0$ in the superconducting phase. Unless otherwise stated, the data are for $J=-U/4$.

\subsection{Relation between gap and self-energy}
\label{sec_gap}

\begin{figure}[ht]
\includegraphics[clip,width=3.4in,angle=0]{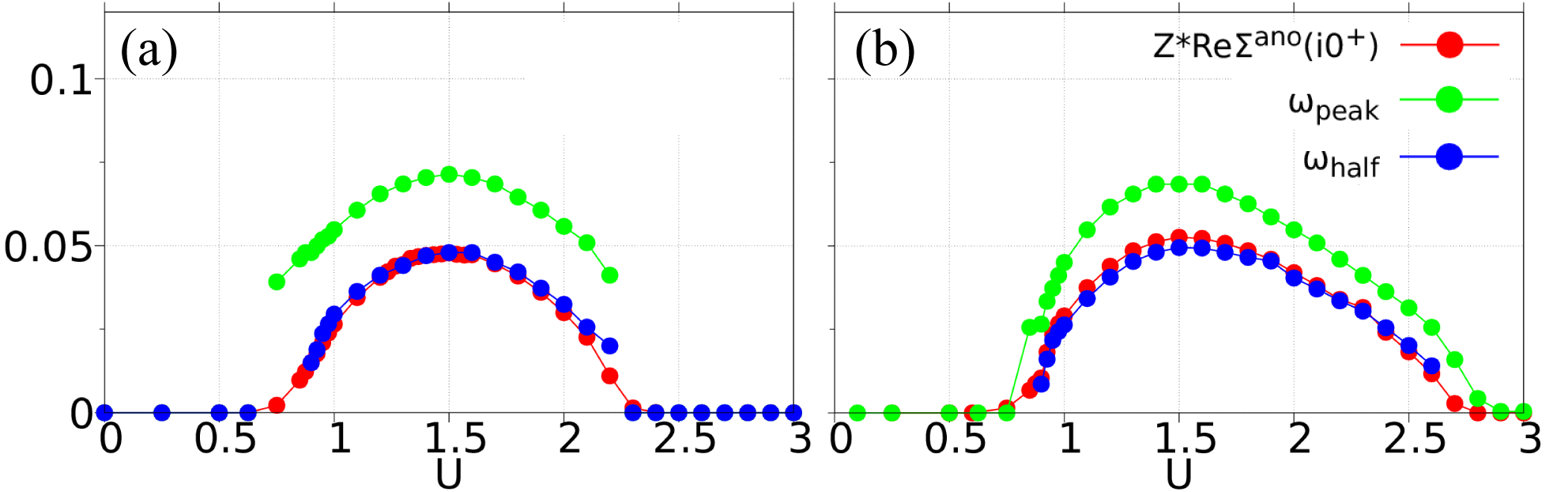}
\caption{(color online) 
The SC gap at $T=0.01$ (panel (a)) and $T=0.005$ (panel (b)) obtained by different procedures. The red dots show 
$Z\mathrm{Re}\Sigma^{\mathrm{ano}}(i0+)$, where $Z$ is the quasi-particle weight.
The green dots represent the low-energy peak positions $\omega_{\text{peak}}$ in the normal fermionic spectral functions $A^{\text{nor}}(\omega)$.
The blue dots are the energies $\omega_{\text{half}}$, where $A^{\text{nor}}$ reaches half the maximum height of 
the peak, i.e. $A^{\text{nor}}(\omega_{\text{half}})=\frac{1}{2}A^{\text{nor}}(\omega_{\text{peak}})$.
}\label{fig:Gap_compare}
\end{figure}

In this subsection, we present the SC gaps obtained by different methods. The SC gap given by the peak energy of $A^\text{nor}(\omega)$ \cite{Gull_prl2013_SC_2D}
generally overestimates the SC gap. When the normal self-energy exhibits a Fermi liquid behavior, $Z\text{Re}\Sigma^{\text{ano}}(i0^+)$ provides a more accurate estimation, which matches the gap value measured at the half-maximum of the peak in $A^\text{nor}(\omega)$, see Fig.~\ref{fig:Gap_compare}.

At very low-temperature, the normal fermionic self-energy 
in the SC phase indeed shows a Fermi-liquid (FL) behavior, i.e. $\Sigma^{\text{nor}}(\omega)\approx (1-Z^{-1})\omega+iO(\omega^2)$. The anomalous self-energy 
scales as $\Sigma^{\text{ano}}(\omega)\approx\text{Re}\Sigma^{\text{ano}}(i0^+)+O(\omega^2)$. 
If we neglect the second order corrections, and insert this FL self-energy into the expression for the 
lattice Green's function, we get

\begin{widetext}
\begin{align}
&\mathbf{G}(\epsilon_{\boldsymbol{k}},\omega) 
=\left[\begin{array}{cc}
\omega-\epsilon_{\boldsymbol{k}}-(1-Z^{-1})\omega & -\mathrm{Re}\Sigma^\text{ano}(i0^{+})\\
-\mathrm{Re}\Sigma^\text{ano}(i0^{+}) & \omega+\epsilon_{-\boldsymbol{k}}-(1-Z^{-1})\omega
\end{array}\right]^{-1} 
=Z\left\{ \omega\sigma_{0}-\left[\begin{array}{cc}
Z\epsilon_{\boldsymbol{k}} & Z\mathrm{Re}\Sigma^\text{ano}(i0^{+})\\
Z\mathrm{Re}\Sigma^\text{ano}(i0^{+}) & -Z\epsilon_{-\boldsymbol{k}}
\end{array}\right]\right\} ^{-1},
\label{eq:Gqpsc}
\end{align}
which is essentially the Green's function of a BCS-type Hamiltonian in Nambu space 
\begin{equation}
H=\sum_{\boldsymbol{k}\nu}
\left[\begin{array}{c} c_{\boldsymbol{k}\uparrow}\\ c_{-\boldsymbol{k}\downarrow}^{\dagger} \end{array}\right]^\dagger
\left[\begin{array}{cc}
Z\epsilon_{\boldsymbol{k}} & Z\mathrm{Re}\Sigma^\text{ano}(i0^{+})\\
Z\mathrm{Re}\Sigma^\text{ano}(i0^{+}) & -Z\epsilon_{-\boldsymbol{k}}
\end{array}\right]
\left[\begin{array}{c} c_{\boldsymbol{k}\uparrow}\\ c_{-\boldsymbol{k}\downarrow}^{\dagger} \end{array}\right].
\label{eq:Heff_qpsc}
\end{equation}
\end{widetext}
Hence, $Z\mathrm{Re}\Sigma^\text{ano}(i0^{+})$ represents the BCS gap or half the total single-particle gap of the normal 
Green's function $G_{11}(\omega)$. Interpreting (\ref{eq:Heff_qpsc}) as the mean-field Hamiltonian of a Hubbard model with bandwidth $2ZD$ and interaction $ZU$, this quantity is related to the interaction by $Z\mathrm{Re}\Sigma^\text{ano}(i0^{+})=ZU\Delta$, with $\Delta$ the local order parameter.  $\mathrm{Re}\Sigma^\text{ano}(i0^{+})/\Delta$ hence represents the effective interaction.

\begin{figure}[t]
\includegraphics[clip,width=\columnwidth,angle=0]{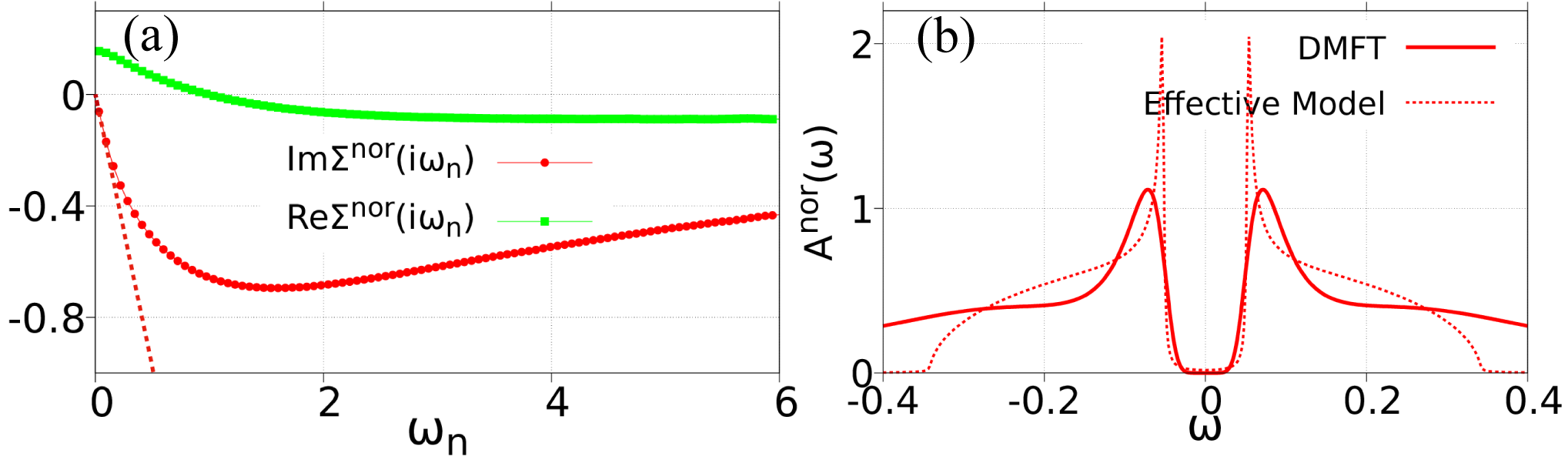}
\caption{(color online)
Matsubara self-energy (panel (a)) and normal Fermionic spectral function (panel (b)) at $T=0.01$, $U=1.5$.
The red dashed line in panel (a) is $y(i\omega_n)=(1-Z^{-1})(i\omega_n)$ with $Z\approx0.34$.
The green dots at $\omega_n\approx 0$ exhibit a parabolic behavior.
In panel (b), the red solid line is the single-particle fermionic spectral function of the three-orbital model, while the red dashed line shows the normal spectral function of the toy model (Eq.~(\ref{eq:Gqpsc})), with a sharp peak at
$\omega=Z\text{Re}\Sigma^{\text{ano}}(i0^+)$  (the bandwidth is renormalized by $Z$). 
The half-maximum position of the low-energy peak in $A^{\text{nor}}$
gives a good estimate of the SC gap.
}
\label{fig:Aw_3band_vs_qpsc}
\end{figure}

Alternatively, we can determine the SC gap from the gap function. In the Matsubara frequency domain, the gap function is defined as
\begin{equation}
\Delta_g(i\omega_n)=\frac{\Sigma^\text{ano}(i\omega_{n})}{1-\frac{\Sigma_{o}(i\omega_{n})}{i\omega_{n}}},
\end{equation}
where $\Sigma_{o}(i\omega_{n})=\frac{\Sigma^\text{nor}(i\omega_{n})-\Sigma^\text{nor}(-i\omega_{n})}{2}=i\mathrm{Im}\Sigma^\text{nor}(i\omega_{n})$ in the presence 
of particle-hole symmetry. In a FL regime, $\Sigma^\text{nor}(i\omega_n)\approx(1-Z^{-1})i\omega_{n}$ and the low-frequency approximation to 
the anomalous self-energy reads $\Sigma^\text{ano}(i\omega_{n})=\mathrm{Re}\Sigma^\text{ano}(i0^{+})$. In this case the gap function is simply a real constant
\begin{equation}
\Delta_{g}(i\omega_{n})=\frac{\mathrm{Re}\Sigma^\text{ano}(i0^{+})}{1-(1-Z^{-1})}=Z\mathrm{Re}\Sigma^\text{ano}(i0^{+})=\Delta_g(\omega).
\end{equation}
The SC gap is determined from the crossing point $\mathrm{Re}\Delta_g(\omega)=\omega$, and we obtain again
$Z\mathrm{Re}\Sigma^\text{ano}(i0^{+})$. 
A comparison between the spectral function of the three-orbital model in the SC phase obtained by means of DMFT and that of the low-energy effective model (\ref{eq:Heff_qpsc}) is shown Fig.~\ref{fig:Aw_3band_vs_qpsc}(b).

\subsection{Orbital-orbital correlation function}

\begin{figure}[t]
\includegraphics[clip,width=3.4in,angle=0]{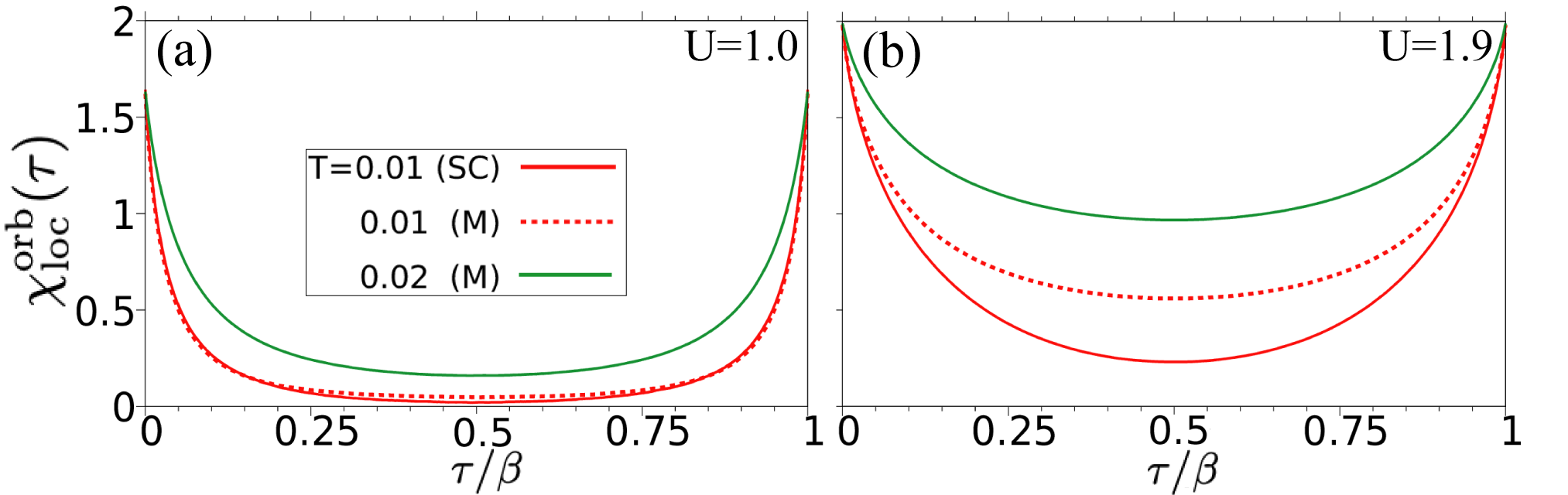}
\caption{(color online) Local imaginary-time orbital-orbital correlation function $\chi^{\mathrm{orb}}_{\mathrm{loc}}(\tau)$. 
Panel (a) is for $U=1.0$ and panel (b) for $U=1.9$. The red solid (dashed) lines are for 
the SC (normal metal) phase at $T=0.01$, and the green solid lines for the normal phase at $T=0.02$. 
}\label{fig:chi_tau_2T}
\end{figure}

Figure~\ref{fig:chi_tau_2T} shows the orbital-orbital correlation function $\chi^{\mathrm{orb}}_{\mathrm{loc}}(\tau)$
in both the SC (solid lines) and normal phase (dashed lines). On the weak-$U$ side of the SC dome ($U=1.0$, panel (a)), 
$\chi^{\mathrm{orb}}_{\mathrm{loc}}(\tau)$ in the SC phase differs only slightly from the result in the normal phase.
On the strong-$U$ side of the dome, however, the orbital freezing leads to a large value of $\chi^{\mathrm{orb}}_{\mathrm{loc}}(\tau)$ near $\tau=\beta/2$ in the metallic state. The transition into the SC state induces particle fluctuations, and hence orbital fluctuations, which results in a more pronounced decay of the correlation function, and hence a larger $\Delta \chi^{\mathrm{orb}}_{\mathrm{loc}}$. This change in the long-time behavior of $\chi^{\mathrm{orb}}_{\mathrm{loc}}(\tau)$ is the direct evidence for the ``unfreezing" of orbital correlations. 

\subsection{$A^{\text{ano}}(\omega)$ and $\text{Im}\chi^{\text{orb}}_{\text{loc}}(\omega)/\pi$}

\begin{figure}[t]
\includegraphics[clip,width=2.4in,angle=0]{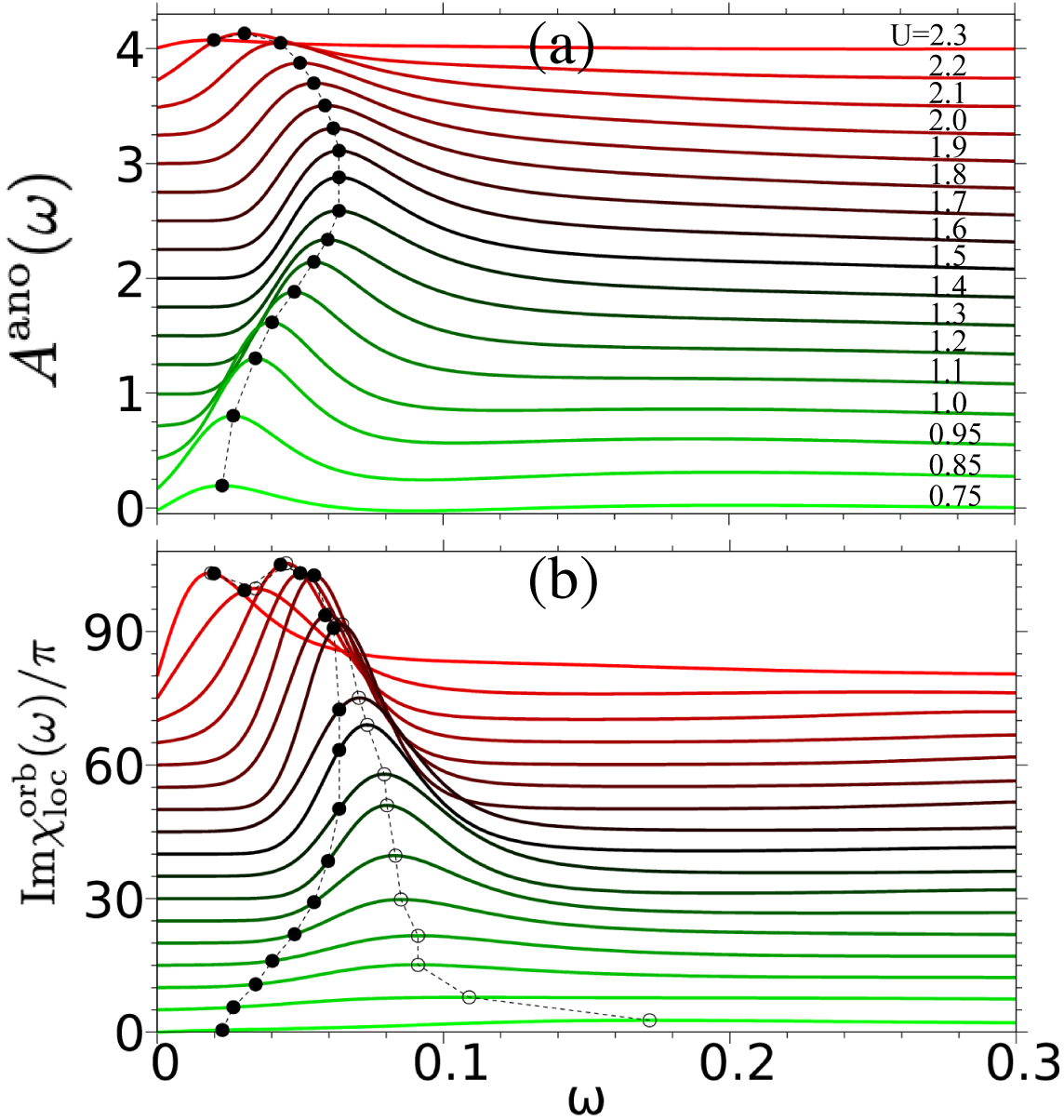}
\caption{(color online) 
Anomalous spectral functions $A^{\text{ano}}(\omega)$ (panel (a)) and bosonic spectral functions 
$\text{Im}\chi^{\text{orb}}_{\text{loc}}(\omega)/\pi$ (panel (b)) for the indicated values of $U$ at $T=0.01$ in the SC phase. The curves in panel (a) [(b)] are shifted by multiples of 0.25 (5.0) along the y-axis for a better presentation.
The full black circles in panel (a) mark the peak positions in $A^{\text{ano}}(\omega)$, and the open circles in panel (b) mark the  peak positions in $\text{Im}\chi^{\text{orb}}_{\text{loc}}(\omega)/\pi$.
For a better comparison between the peak positions, we also indicate the peak positions from panel (a) in panel (b). 
}\label{fig:Aanow_Chiw_T0p01_Ux}
\end{figure}

As supplementary data to Fig.~\ref{fig:peaks_UxTx}(a), we show in Fig.~\ref{fig:Aanow_Chiw_T0p01_Ux} the evolution of the anomalous spectra $A^{\text{ano}}(\omega)$ and the  
bosonic spectra $\text{Im}\chi^{\text{orb}}_{\text{loc}}(\omega)/\pi$ for $T=0.01$ in the $U$ range from $U=0.75$ to $U=2.3$, where a SC solution exists. For a better comparison of the peak positions, we indicate the peak energies in $A^{\text{ano}}(\omega)$ from panel (a) by the full black dots in panel (b), while the open dots in panel (b) indicate the peak positions for $\text{Im}\chi^{\text{orb}}_{\text{loc}}(\omega)/\pi$. As one can see, the two peak energies merge on the strong-$U$ side of the SC dome.

\subsection{Spectral functions for smaller $|J/U|$}

\begin{figure}[t]
\includegraphics[clip,width=2.2in,angle=0]{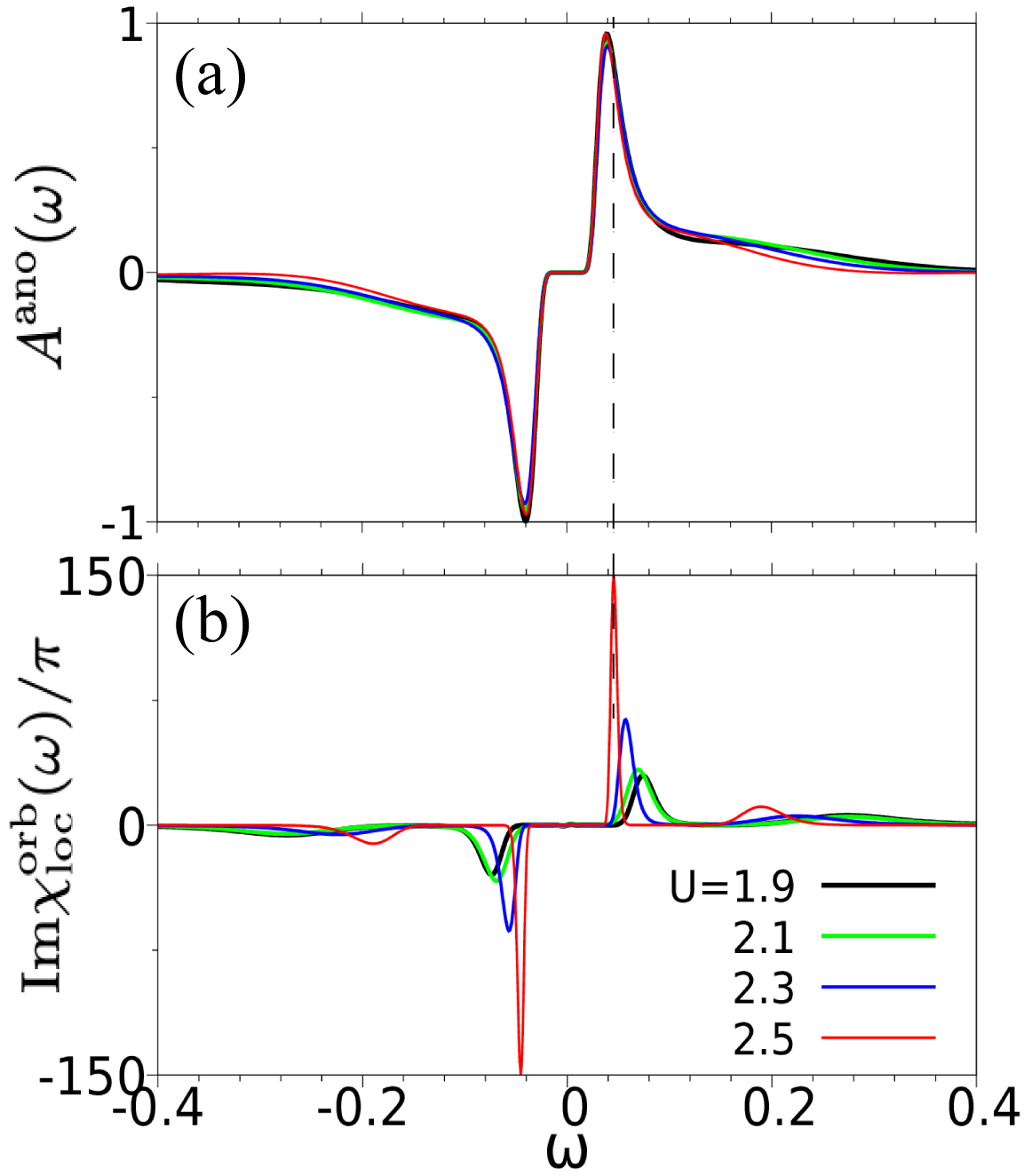}
\caption{(color online) Anomalous spectral function $A^{\mathrm{ano}}(\omega)$ (panel (a)), 
and bosonic spectrum $\tfrac1\pi\mathrm{Im}\chi^{\mathrm{orb}}_{\mathrm{loc}}(\omega)$ (panel (b)) for the indicated values of $U$, $J/U\approx-0.13$ and $T=0.005$. 
}\label{fig:AwChiw_JdivU0p13_UxT0p005}
\end{figure}

For smaller $|J/U|$, the maximum $T_c$ is reduced and the Mott transition line shifts to larger $U$. Hence, we need to treat lower $T$ and larger $U$ to reach the large-$U$ regime of the SC phase. 
Figure~\ref{fig:AwChiw_JdivU0p13_UxT0p005} shows the evolution of $A^{\mathrm{ano}}(\omega)$
and $\tfrac1\pi\mathrm{Im}\chi^{\mathrm{orb}}_{\mathrm{loc}}(\omega)$ as one increases the interaction $U$ for $|J/U|\approx0.13$ at $T=0.005$. At $U=1.9$, the peak positions in $A^{\mathrm{ano}}(\omega)$ and $\tfrac1\pi\mathrm{Im}\chi^{\mathrm{orb}}_{\mathrm{loc}}(\omega)$ are still well separated, since this parameter set is now on the weak-pairing side of the dome. While the peak positions in $A^{\mathrm{ano}}(\omega)$ change only slightly as $U$ increases, the peak energies for $\tfrac1\pi\mathrm{Im}\chi^{\mathrm{orb}}_{\mathrm{loc}}(\omega)$
rapidly shift to lower values. At $U=2.5$, the peaks in the fermionic and bosonic spectral functions nearly match (lock-in phenomenon).

\begin{figure}[t]
\includegraphics[clip,width=2.4in,angle=0]{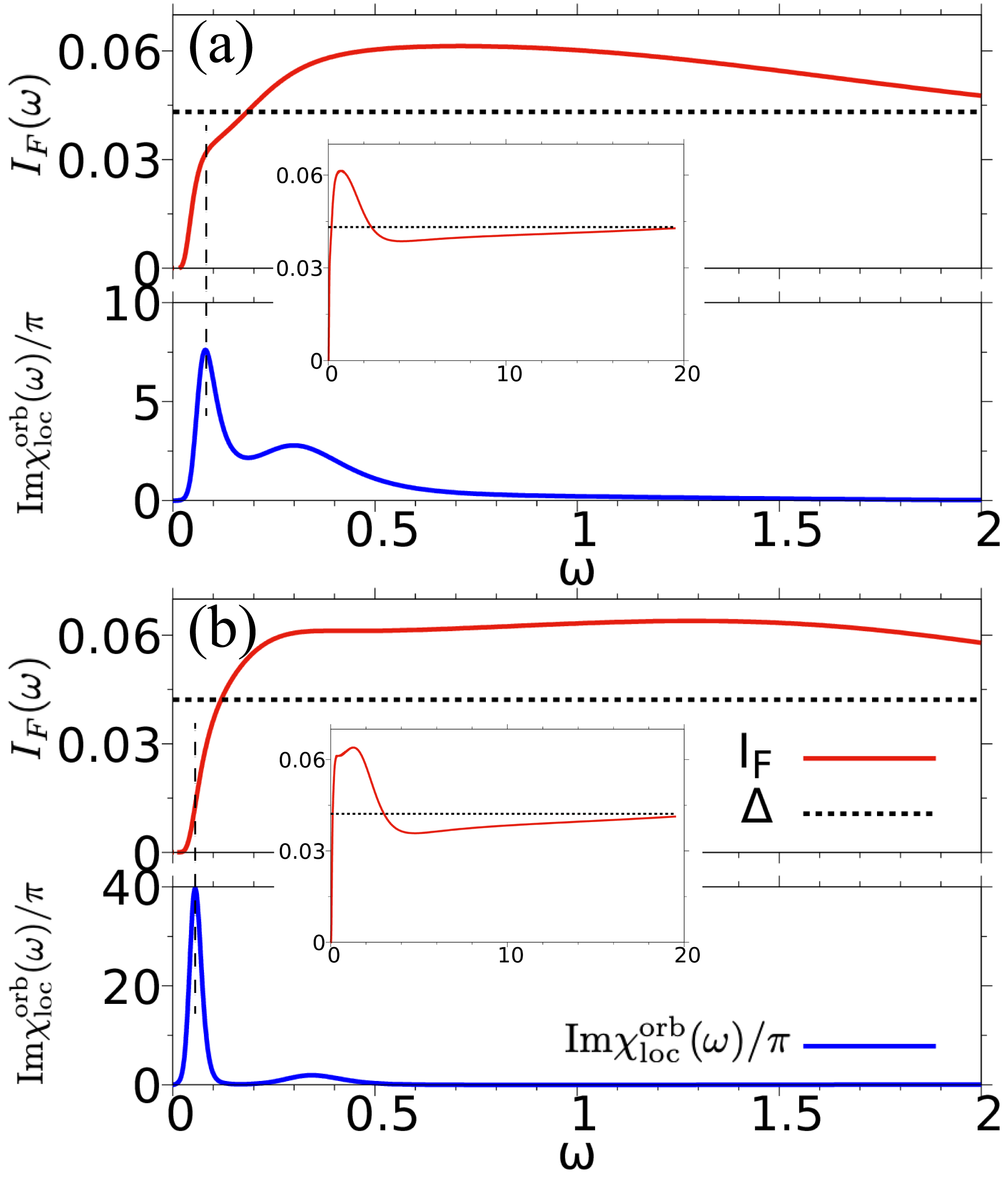}
\caption{(color online) Cumulative spectral weight $I_F(\omega)$ (upper panel) and
$\mathrm{Im}\chi_{\text{loc}}^{\mathrm{orb}}(\omega)$ (lower panel) for $U=1.0$ (panel (a)) and $U=1.9$ (panel (b)) at $T=0.01$.
The dashed vertical line in panel (b) links the low energy bosonic peak in 
$\mathrm{Im}\chi_{\text{loc}}^{\mathrm{orb}}(\omega)$ to the fast increase 
in $I_F(\omega)$. The horizontal dashed line in the upper panel is the SC order parameter $\Delta$. The insets
 show $I_F(\omega)$ in the frequency range $\omega\in[0,20]$, illustrating the limit $\Delta=\lim_{\omega\rightarrow\infty}I_{F}(\omega)$. 
}\label{fig:IFw_Chiw}
\end{figure}

\subsection{Cumulative Spectral Weight of the SC Order Parameter}
The SC order parameter is related to the anomalous Green's function and spectral function by
\begin{align}
\Delta= &-F(\tau=0^{+}) 
= &-\frac{1}{\pi}\int_{-\infty}^{+\infty}\mathrm{d\omega}\mathrm{Im}F(\omega)\frac{1}{1+e^{-\beta\omega}}.
\end{align}
Using the property $\text{Im}F(\omega)=-\text{Im}F(-\omega)$ we have
$\Delta =-\frac{1}{\pi}\int_{0}^{+\infty}\mathrm{d\omega}\mathrm{Im}F(\omega)\frac{e^{\beta\omega}-1}{e^{\beta\omega}+1}.$
Thus one may define the cumulative spectral weight contribution to $\Delta=\lim_{\omega \rightarrow\infty}I_{F}(\omega)$ as
$I_{F}(\omega^\prime)=-\frac{1}{\pi}\int_{0}^{\omega^\prime}\mathrm{d\omega}\mathrm{Im}F(\omega)\frac{e^{\beta\omega}-1}{e^{\beta\omega}+1}$.
In the low-temperature limit $\beta=1/T\rightarrow\infty$,  
$\frac{e^{\beta\omega}-1}{e^{\beta\omega}+1}\rightarrow 1$, and one finds
\begin{equation}
\lim_{T\rightarrow0}I_{F}(\omega^\prime)=-\frac{1}{\pi}\int_{0}^{\omega^\prime}\mathrm{d\omega}\mathrm{Im}F(\omega),
\end{equation}
which is identical to Eq.~(2) in Ref.~\onlinecite{Kyung2009}.

The cumulative spectral weight for the system at $T=0.01$, $U=1.9$ is shown in Fig.~\ref{fig:IFw_Chiw}(b). 
The match between the peaks in $\text{Im}F(\omega)$ and $\mathrm{Im}\chi_{\text{loc}}^{\mathrm{orb}}(\omega)$
yields a direct correspondence between the fast increase in $I_F(\omega)$ and the peak in the bosonic spectrum.
In Fig.~\ref{fig:IFw_Chiw}(a), we also show the data for $U=1.0$, where the peak in 
$\text{Im}F(\omega)$ and $\mathrm{Im}\chi_{\text{loc}}^{\mathrm{orb}}(\omega)$ appears at a higher  
energy than the fast increase in $I_F(\omega)$.

The ``overshooting" of $I_F(\omega)$ is similar to what has been reported in Ref.~\onlinecite{Kyung2009} for the under-doped Hubbard model. 

\subsection{$E_r$ versus SC gap}
\begin{figure}[b]
\includegraphics[clip,width=\columnwidth,angle=0]{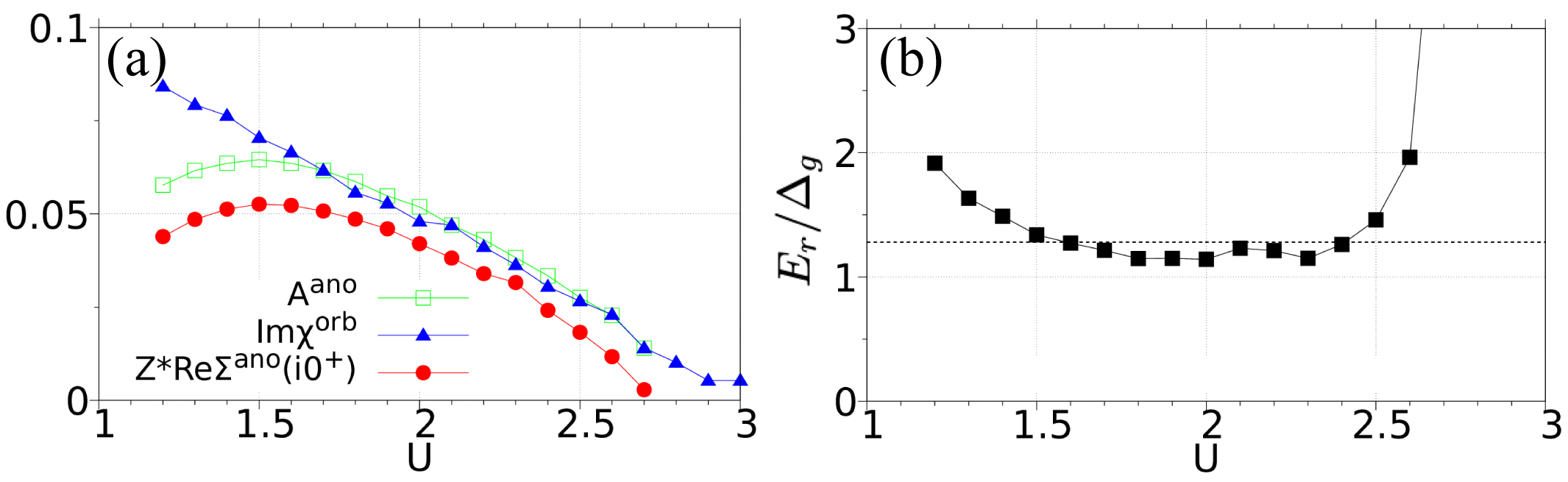}
\caption{(color online) 
Panel (a): The low-energy peak positions in $\mathrm{Im}F(\omega)$, $\mathrm{Im}\chi_{\text{loc}}^{\mathrm{orb}}(\omega)$ (denoted as $E_r$) and the SC gap $\Delta_g=Z \text{Re}\Sigma^{\text{ano}}(i0^+)$
as a function of $U$ at $T=0.005$, respectively. Panel (b): The ratio $E_r/\Delta_g$ compared to the experimentally found value (dashed line). 
}\label{fig:E_r_div_SCgap}
\end{figure}

Figure~\ref{fig:E_r_div_SCgap} plots the evolution of the low-energy peak positions in $\mathrm{Im}F(\omega)$, the low-energy peak positions in $\mathrm{Im}\chi_{\text{loc}}^{\mathrm{orb}}(\omega)$ ($\equiv E_r$) and the SC gap ($\Delta_g$) as a function of $U$ at
$T=0.005$, respectively. On the strong-$U$ side, the peak positions in $\mathrm{Im}F(\omega)$ (green) overlap with those of $\mathrm{Im}\chi_{\text{loc}}^{\mathrm{orb}}(\omega)$ (blue). 
The ratio $E_r/\Delta_g\approx 1.2$ for $1.6<U<2.4$ is consistent with the experimentally found value ($\approx 1.28$) in a broad range of unconventional SCs \cite{NatPhys2009_Ereson_2SCgap}. 

\end{document}